\documentclass[a4paper,12pt]{article}
\usepackage[T2A]{fontenc}			
\usepackage[utf8]{inputenc}         
\usepackage{indentfirst}             
\usepackage[warn]{mathtext}          
\usepackage[english]{babel} 
\usepackage{misccorr}                
\usepackage{verbatim}                
\usepackage[dvips]{graphicx}         
\usepackage{multicol}                
\usepackage{mathrsfs,amsmath,amssymb,amsbsy} 
\usepackage[left=1.5cm,right=1cm,
top=1cm,bottom=1cm,bindingoffset=0cm]{geometry}
\usepackage[cm]{fullpage}
\usepackage{graphicx}
\usepackage{xcolor}
\def\beq{\begin{equation}}
\def\eeq{\end{equation}}
\def\bear{\begin{eqnarray}}
\def\bearr{\begin{eqnarray} \lal}
\def\ear{\end{eqnarray}}
\def\earn{\nonumber \end{eqnarray}}
\def\bear{\begin{eqnarray}}
\def\bearr{\begin{eqnarray} \lal}
\def\lal{&&\nqq {}}

\def\d{\partial}

\def\mn{_{\mu\nu}}

\begin{document}
	
	\begin{center}
		\textbf{Cosmological solutions of a chiral self-gravitating model of $f(R,(\nabla R)^2,\square R)$ gravity}
	\end{center}
	\begin{center}
		\textbf{T.I. Chaadaeva$^*$, S.V. Chervon$^*$$^\dag$$^\star$}
	\end{center}
	\begin{center}
		\textit{$^*$Laboratory of gravitation, cosmology, astrophysics,\\ Ulyanovsk State Pedagogical University, \\ Lenin's square, 4/5, Ulyanovsk, 432071, Russia}\\[0.2cm]
	\end{center}
	\begin{center}
		\textit{$^\dag$Physics Department, \\Bauman Moscow State Technical University, \\2-nd Baumanskaya street 5, Moscow, 105005, Russia}\\[0.2cm]
	\end{center}
	\begin{center}
		\textit{$^\star$Institute of Physics, \\Kazan Federal University, \\Kremlevskaya street 18, Kazan, 420008, Russia}\\[0.2cm]
	\end{center}
	
	\textbf{Abstract}
	
	\noindent
	
We study modified $f(R,(\nabla R)^2,\square R)$ gravity and show in detail how it can be reduced to Einstein gravity with a few scalar fields and then represented in the form of chiral self-gravitating model of the special type. In further investigation of the model we focus on cosmology and looking for solutions of the dynamic equations of chiral fields and the Einstein-Friedman equations in the Friedman-Robertson-Walker spacetime. Exact solutions of the considered model for zero and constant potential are found. Between them power law solution corresponded to equation of state for radiation dominated Universe, de Sitter solution, trigonometric and hyperbolic expansion solution. Various type of chiral fields evolution support listed above solutions.

	
	
\section{Introduction}
	
Modified theories of gravity began to develop especially actively after experimental confirmation of the accelerated expansion of the Universe, when it became obvious that there was a shortage of tools of general relativity to describe the experimental data obtained. One of the well-known theories is the $f(R)$ theory of gravity, which is able to describe a number of phenomena (not only the accelerated expansion of the universe, but also inflation at an early and late stage) without introducing additional fields (dark energy, dark matter). The modification of the theory consists in the introduction into the integral of the action of the function of scalar curvature $R$. Replacing $f(R) = R$ allows us to obtain the theory of GR. Subsequently, attempts were made to consider various versions of the function $f(R)$. The most successful and consistent with the observational data is the model of A. Starobinsky $f(R) = R+ \frac{R^2}{6M^2}$, where the value $M$ has the dimension of mass. In particular, the Starobinsky model allows us to explain the modern acceleration of the Universe, inflation.

Higher-order curvature corrections to the Einstein-Hilbert effect occur when quantum effects are considered in the low-energy limit of string theory, superstrings, and supergravity, necessary for the construction of a quantum theory of gravity. An example of the application of quantum corrections was demonstrated by A. Starobinsky in cosmology. It has been shown that such corrections can control the accelerated expansion of the universe at an early stage of its evolution (inflation). Similar models were developed taking into account 6th-order corrections in gravity theories of the form $R+\alpha R^2+\gamma R\Box R$ \cite{gottloeber90},
%
where $\alpha$, $\gamma$ are some constants.
$R^2$, $R\Box R$ are additional terms modifying Einstein's theory, which can be replaced by two scalar fields using conformal transformations of the metric, leading to general relativity with two scalar fields.

In \cite{Naruko:2015zze}, a model of gravity is considered, where the function $f$ contains derivatives of  high orders on scalar curvature (till the second one). In the same work, a method of transition to GR with scalar fields is indicated. The developed method has been successfully applied on several models of $f(R)$ gravity with higher derivatives.
For example, in the work \cite{CHNM2017} the derivation of equations for $f(R,(\nabla R)^2) = f_1(R) + X(R)R_{,\mu}R^{,\nu}$ model represented in detail. Further investigation with consideration of cosmological aspects and connection the model with scalar tensor representation are developed in \cite{Chervon:2018ths}. Cosmological solutions consistent with observational data are studied in the work
\cite{Chervon:2019jfu}.

The truncated model $f(R,\Box R)$, reduced to the chiral cosmological model with three scalar fields, is presented in \cite{ChFCh20231}, \cite{ChFCh20232}. The set of cosmological exact solutions contains  radiation dominated expansion, hyperbolic and trigonometric evolution of the scale factor.  Slow roll solution leads to de Sitter expansion with one dynamical field. Also inclusion of additional material field into consideration leads to new classes of exact solution with scalar fields and perfect fluid. To make possible comparison with observation data the way of construction one-field cosmological model for this purpose proposed in \cite{ChFCh20232}.

In this paper, we consider a modified theory of gravity of the form $f(R, (\nabla R)^2,\Box R)$ and some cosmological solutions.
Let us stress now that our main target is to study cosmological multi-field models in E-frame originated from gravity theory with higher derivatives, i.e. chiral cosmological models where the potential and chiral metric dictated by $f(R, (\nabla R)^2,\Box R)$ model.

Material of the article is presented as follow.
In Sec. 2, the method of translating the model to Einstein gravity with a few scalar fields and then represented it in the form of chiral self-gravitating model of the special type is described in detail.
It is also indicated that the model written in the Einstein frame can be written as a three-component effective chiral cosmological model. The dynamic equations of the model are presented in Sec. 3. Cosmological solutions based on the choice of zero potential are described in Sec. 4. Sec. 5 contains solutions based on the choice of constant potential. The final section provides some conclusions of the work.

\section{The action for $f(R,(\nabla R)^2,\square R)$ gravity in Einstein frame}

In the	present work we study the general case with the action
\beq \label{act-g1}
S = \int d^4x\sqrt{-g}\left[f(R,(\nabla R)^2,\square R )\right],
\eeq
where $R$ is a scalar curvature, $(\nabla R)^2=\nabla_\mu R\nabla^\mu R$, $\square R= \nabla_\mu \nabla^\mu R$.

	Following the method represented in the paper \cite{Naruko:2015zze} we can transform the model \eqref{act-g1} to Einstein gravity with scalar fields.
To get this target
we introduce the lagrangian multipliers $\tilde{\lambda},\tilde{\Lambda}_1, \tilde{\Lambda}_2$  with the corresponding additional fields
$\phi, X, B$.
	Thus, the action \eqref{act-g1} is transformed to
\beq           \label{act-phiB}
		S=\int d^4x \sqrt{-g} \left[f(\phi,X,B) - \tilde{\lambda}(\phi-R)-\tilde{\Lambda}_1(X-(\nabla R)^2) -\tilde{\Lambda}_2(B-\square R) \right].
\eeq
	
Let us note that fields $\phi, X, B$ are independent. Our target is to connect them by the following way $\phi=R,~~X=g^{\mu\nu}\nabla_{\mu}\phi\nabla_{\nu}\phi,~~B=\Box \phi $.
	The variation of the action \eqref{act-phiB} with respect to the fields leads to the equations
	\begin{eqnarray}
	F_1 \equiv	\frac{\d      f}{\d      \phi}- \frac{\d      \tilde{\lambda}}{\d      \phi}(\phi-R)-  \tilde{\lambda}- \frac{\d      \tilde{\Lambda}_1}{\d      \phi}(X-(\nabla R)^2) -\frac{\d      \tilde{\Lambda}_2}{\d      \phi} (B-\square R)  &=& 0, \\
F_2	\equiv	\frac{\d      f}{\d      X}- \frac{\d      \tilde{\lambda}}{\d      X}(\phi-R)-  \frac{\d      \tilde{\Lambda}_1}{\d      X}(X-(\nabla R)^2)-\tilde{\Lambda_1} -\frac{\d      \tilde{\Lambda}_2}{\d      X} (B-\square R)  &=& 0,\\
F_3	\equiv	\frac{\d      f}{\d      B}- \frac{\d      \tilde{\lambda}}{\d      B}(\phi-R)- \frac{\d      \tilde{\Lambda}_1}{\d      B}(X-(\nabla R)^2)- \frac{\d      \tilde{\Lambda}_2}{\d      B} (B-\square R)-  \tilde{\Lambda}_2&=& 0.
	\end{eqnarray}

Combination $F_1 d\phi + F_2 dX +F_3 dB$	leads to the following equation
\begin{equation}
df=(\phi-R)d\tilde{\lambda}+(X-(\nabla R)^2)d\tilde{\Lambda}_1 +(B-\square R)d\tilde{\Lambda}_2+\tilde{\lambda}d\phi +\tilde{\Lambda}_1 dX +\tilde{\Lambda}_2 dB.
\end{equation}

Thus we could not set $ \tilde{\lambda},\tilde{\Lambda}_1,\tilde{\Lambda}_2$ equal to constants. In opposite case we obtain the restriction on the form of functional dependence of $f: f=\tilde{\lambda}\phi +\tilde{\Lambda}_1 X +\tilde{\Lambda}_2 B$. Substitution of this form back to the action \eqref{act-g1} leads to a special case of a model with the action
\begin{equation}
S=\int d^4x \sqrt{-g} \left[\tilde{\lambda}R+\tilde{\Lambda}_1(\nabla R)^2+\tilde{\Lambda}_2\square R \right].
\end{equation}
Therefore we can state that the lagrangian multipliers are determined dynamically.

If we variate the action \eqref{act-g1} by lagrangian multipliers
$ \tilde{\lambda},\tilde{\Lambda}_1,\tilde{\Lambda}_2$ we get the relations
\begin{equation}
\phi=R,~~X=g^{\mu\nu}\nabla_{\mu}R\nabla_{\nu}R,~~B=\Box R.
\end{equation}

Next step will be to find the way of transforming
%
the multipliers $(\tilde{\lambda},\tilde{\Lambda}_1,\tilde{\Lambda}_2)$ to $(\lambda, \Lambda_1,\Lambda_2)$, to get the constraint equations instead of dynamical equations for them. The transformation \cite{Naruko:2015zze} is
\beq           \label{tl-l}
		\lambda=\tilde{\lambda}-\nabla^\mu\left[\tilde{\Lambda}_1\nabla_\mu (\phi +R)\right]-\square \tilde{\Lambda}_2,~~\Lambda_1=\tilde{\Lambda}_1 ~~\Lambda_2=\tilde{\Lambda}_2.
\eeq

Now, using \eqref{tl-l}, we can insert $(\tilde{\lambda},\tilde{\Lambda}_1,\tilde{\Lambda}_2)$ expressed over $(\lambda, \Lambda_1,\Lambda_2)$ in \eqref{act-phiB} and analyse the result.
	
The $\Lambda_1$-term in Lagrangian
$$
\nabla^\mu\left[\tilde{\Lambda}_1\nabla_\mu (\phi +R)\right](\phi-R)
$$
	we can transform by the following way.
	Considering 4-divergence
$$
\nabla^\mu\left[\left[\tilde{\Lambda}_1\nabla_\mu (\phi +R)\right] (\phi-R)\right]=\nabla^\mu\left[\tilde{\Lambda}_1\nabla_\mu (\phi +R)\right](\phi-R)+\tilde{\Lambda}_1 [(\nabla \phi)^2-(\nabla R)^2],
$$
we can exchange in the Lagrangian the term $\nabla^\mu\left[\tilde{\Lambda}_1\nabla_\mu (\phi +R)\right](\phi-R) $
on $-\tilde{\Lambda}_1 [(\nabla \phi)^2-(\nabla R)^2] $ taking into account vanishing of the 4-divergence by Gauss-Stokes theorem.
After that we simplify $\Lambda_1$-part of the Lagrangian
$$
-\tilde{\Lambda}_1(X-(\nabla R)^2) +\tilde{\Lambda}_1 [(\nabla \phi)^2-(\nabla R)^2]=-\tilde{\Lambda}_1(X-(\nabla \phi)^2).
$$
	
	Similarly the $\Lambda_2$-term
	$$
	\square \tilde{\Lambda}_2(\phi-R)
	$$
	we can represent as
	$$
	\square \tilde{\Lambda}_2(\phi-R)=\nabla^\mu M_\mu - \tilde{\Lambda}_2\left(\square \phi-\square R\right)
	$$
	where
	$$
	M_\mu= \nabla_\mu\left[\tilde{\Lambda}_2 (\phi-R)\right].
	$$
	
	After that we transform  $\Lambda_2$-part of the Lagrangian to
	$$
	\tilde{\Lambda}_2\left(\square \phi-\square R\right)-\tilde{\Lambda}_2\left(B-\square R\right)=-\tilde{\Lambda}_2\left(B-\square \phi\right).
	$$
	
	Finally we get the action
\beq           \label{act-4}
		S=\int d^4x \sqrt{-g} \left[f(\phi,X,B) - \lambda(\phi-R)
		-\Lambda_1 (X-(\nabla \phi)^2)- \Lambda_2 (B-\square \phi) \right].
\eeq

Since the action does not include any derivatives terms of $X$ and $B$, variation of the action wrt those variables yield constraint equations
$$
\frac{\d      f}{\d      X}=\Lambda_1,~~\frac{\d      f}{\d      B}=\Lambda_2
$$
rather then dynamical equations of motion, which can be plugged back into the action without changing the nature of the theory.

	For the action \eqref{act-4} variation with respect to $(\Lambda_1, \Lambda_2)$ leads to the constraint equations
	\begin{eqnarray}
		\nonumber
		X &=& (\nabla \phi)^2,\\
		\label{B-f}
		B &=&\square \phi.
	\end{eqnarray}
	
	Obtained constraints can be plugged back into the action \eqref{act-4} without changing the nature of the theory.
	Thus, the action takes the form
\beq           \label{act-5}
		S=\int d^4x \sqrt{-g} \left[f(\phi,(\nabla \phi)^2),\square \phi) - \lambda(\phi-R) \right].
\eeq

Note that here in conversed action in J-frame we have the function $f$ which depends on the scalar field $\phi$ and its first and second derivative.
		
%
The Lagrange multiplier can be introduced by replacing $R$ with $\phi$, all derivatives of $R$ are replaced by derivatives of $\phi$ as well. So, it is possible to determine the dynamics of $\phi$ by terms reflecting the derivative of $\phi$, and by varying the action by $\phi$, then $\lambda$ is a dynamic field.

Let us transform the model \eqref{act-5} in Jordan frame (J-frame) to Einstein frame (E-frame) using conformal transformation $g^E_{\mu\nu}=\Omega^2(x)g^J_{\mu\nu}, ~\Omega^2(x)=2\lambda (x) $. As a result we get
\begin{equation}\label{act-br}
S=\int d^4x \sqrt{-g} \left[\frac{1}{2} R-\frac{1}{2}\left(\frac{3}{2\lambda^2} g^{\mu\nu} \lambda_{,\mu}\lambda_{,\nu}\right) + \frac{1}{4\lambda^2}f(\phi,(\nabla \phi)^2),\square \phi) - \frac{\phi}{4\lambda} \right].
\end{equation}

It is clear that values in \eqref{act-br} are belonging to E-frame (we omit the index "E" over the metric $g$ and scalar curvature $R$).

\subsection{Model with reducing order}	

We reduce the order of derivatives introduced by the $\square\phi$ term by introducing the Lagrange multiplier and the auxiliary field associated with it:

\beq           \label{act-4-1}
		S=\int d^4x \sqrt{-g} \left[f(\phi,(\nabla \phi)^2,B) - \lambda(\phi-R)
		- \Lambda (B-\square \phi) \right]
\eeq
	
	Variation of the action wrt $B$ gives a constraint equation
\beq           \label{2-6}
		\Lambda=\frac{df}{dB}\equiv f_B.
\eeq
	
The introduced restriction excludes $\Lambda$ from the action. If we assume that
$$
f_{BB}\neq 0,
$$
then $\Lambda$ will appear in action again.

In other words, if $f_B=const.$ we cannot vary $\Lambda$ in \eqref{2-6}.

Note in further discussion we could not set $B=\square \phi$. If we do this we must return back to the action \eqref{act-5} and working in the framework of this model.
	

	Let us note that $ \Lambda \neq const.$ (in the opposite case the relation
	$B=\square \phi$ may not be valid), thus  $f_{BB} \neq 0$.  In that case we can introduce new field
\beq           \label{varp}
		\psi =f_B
\eeq
	and considering $ (g_{\mu\nu}, \lambda, \phi, \psi )$ as the basis system. It is justified since the transformation $ (g_{\mu\nu}, \lambda, \phi, B)$ to
	$ (g_{\mu\nu}, \lambda, \phi, \psi )$  locally reversibly under the condition  $f_{BB} \neq 0$.

Let us consider this transformation in detail. First of all we have to change the arguments of the function $f: f(\phi, (\nabla \phi)^2, B) $ to $ f(\tilde{\phi}, \tilde{(\nabla \phi)}^2, \tilde\psi). $ Jacobian of the transformation
\begin{equation}
\tilde{\phi}=\tilde{\phi}(\phi, (\nabla \phi)^2, B),~~
\tilde{(\nabla \phi)}^2=\tilde{(\nabla \phi)}^2(\phi, (\nabla \phi)^2, B),~~\tilde\psi=\tilde\psi (\phi, (\nabla \phi)^2, B),
\end{equation}
under conditions $ \tilde\phi =\phi,~ \tilde{(\nabla \phi)}^2=(\nabla \phi)^2,~\tilde\psi =\psi$, is equal to $f_{BB}$.
As $f_{BB} \neq 0$ the transformation above is reversible and we can express new arguments as
\begin{equation}
\phi=\phi(\tilde{\phi}, \tilde{(\nabla \phi)}^2, \tilde\psi),~~
(\nabla \phi)^2=(\nabla \phi)^2(\tilde{\phi}, \tilde{(\nabla \phi)}^2, \tilde\psi),~~B=B (\tilde{\phi}, \tilde{(\nabla \phi)}^2, \tilde\psi).
\end{equation}
Using conditions above, we can write
$$
B=B (\phi, (\nabla \phi)^2, \psi).
$$

	Let us consider the action \eqref{act-4-1} again. To form the 4-divergence
\beq           \label{div-2}
		\nabla^\mu (\Lambda \nabla_\mu \phi)=(\nabla^\mu \Lambda) \nabla_\mu \phi
		+ \Lambda \square \phi
\eeq
	we subtract and add the term
	$ (\nabla^\mu \Lambda) \nabla_\mu \phi $ in the action \eqref{act-4-1}.
	Then, avoiding the 4-divergence and using \eqref{2-6}, we get the action
\beq           \label{act-7}
		S=\int d^4x \sqrt{-g} \left[  \lambda R -(\nabla^\mu f_B) \nabla_\mu \phi + f(\phi,(\nabla \phi)^2, \psi) - f_B B(\phi, (\nabla \phi)^2, \psi) -  \lambda \phi \right].
\eeq
	
	Equivalently, in J-frame, we have
\beq           \label{act-7-1}
		S=\int d^4x \sqrt{-g} \left[  \lambda R -(\nabla^\mu \psi) \nabla_\mu \phi
		+ f(\phi,(\nabla \phi)^2,\psi) - \psi B(\phi, (\nabla \phi)^2, \psi) -  \lambda \phi \right].
\eeq
	
  Let us transform the action \eqref{act-7-1} from the Jordan frame with the metric
  $g\mn = g^J\mn$ to the E-frame using the conformal transformation
  $g^E_{\mu\nu}=\Omega^2(x)=2\lambda (x)g^J_{\mu\nu}$
  As a result, in the E-frame we have	
\bear           \label{act-8a}
		S=\int d^4x \sqrt{-g}\left[ \frac{R}{2}-\frac{1}{2}\left(\frac{3}{2}\lambda_\mu \lambda^\mu \right) +\frac{1}{4\lambda^2}f(\phi,(\nabla\phi)^2,\psi)-\frac{1}{4\lambda^2}\psi B(\phi,(\nabla\phi)^2,\psi)- \frac{1}{4\lambda}\phi\right]
\ear
  where we also used transformation of a scalar field $\phi^J=\sqrt{2\lambda}\phi^E $.

  We see that in the general case we cannot present the model with the action \eqref{act-8a} as a
   a chiral self-gravitating model. To achieve this goal, we choose
  the function $B$ by the following way:
\beq           \label{B12}
		B(\phi, (\nabla\phi)^2, \psi) = B_1(\phi,\psi)
					g^{\mu\nu} \phi_{,\mu}\phi_{,\nu}+ B_2(\phi,\psi).
\eeq
  and similarly the function $f$ as	
\beq \label{f24}
		f(\phi, (\nabla\phi)^2, \psi) = f_1(\phi,\psi)g^{\mu\nu} \phi_{,\mu}\phi_{,\nu}
		+ f_2(\phi,\psi).
\eeq

	Substitution \eqref{B12} and \eqref{f24} in the action \eqref{act-8a}, with $\lambda = e^{\sqrt{\frac{2}{3}}\chi}$, leads to
	\begin{eqnarray}\label{act-8b}
\nonumber
		S=\int d^4x \sqrt{-g} \left[\frac{1}{2} R-\frac{1}{2}\left( (\nabla\chi)^2- e^{-\sqrt{\frac{2}{3}}\chi}(\phi_{,\mu}\psi^{,\mu})+ e^{-2\sqrt{\frac{2}{3}}\chi} \left(\psi B_1(\phi,\psi)-\frac{1}{2} f_1(\phi,\psi) \right)
(\nabla\phi)^2\right) - \right. \\
\left. \frac{1}{4} e^{-\sqrt{\frac{2}{3}\chi}}\left(\phi+ e^{-\sqrt{\frac{2}{3}}\chi}\left(\psi B_2(\phi,\psi) - f_2(\phi,\psi)\right)\right)\right].
	\end{eqnarray}

	In such a way we get the action \eqref{act-8b} which can be described by the three-component chiral self-gravitating model with the target space metric with non-zero components
	\begin{equation}\label{ts3d}
		h_{11}=1,~~h_{22}=e^{-2\sqrt{\frac{2}{3}}\chi}\left( \psi B_1(\phi,\psi)-\frac{1}{2}f_1(\phi,\psi)\right),~~h_{23}= -\frac{1}{2}e^{-\sqrt{\frac{2}{3}}\chi}.
	\end{equation}
	
	The potential of the interaction is
	\begin{equation}\label{V-1}
		W(\chi,\phi,\psi)=\frac{1}{4}\left(e^{-\sqrt{\frac{2}{3}}\chi }\phi+ e^{-2\sqrt{\frac{2}{3}}\chi}\left(\psi B_2(\phi,\psi) -f_2(\phi,\psi)\right)\right).
	\end{equation}
	
	
\section{Dynamic equations of the model}
Let us use the general form of the chiral field equations in the FRW metric represented in \cite{Chervon:2019nwq}
\begin{equation}
\label{f-1}
	-h_{CB}\left(\ddot{\phi}^B+3H\dot{\phi}^B\right) - h_{CB,D}\dot{\phi}^D\dot{\phi}^B
+ \frac{1}{2}h_{DB,C}\dot{\phi}^D\dot{\phi}^B - W_{,C} = 0.
\end{equation}

Using our designations for the fields, the metric and the potential we have the following equations of the chiral fields
\begin{equation}\label{chi-1}
-\ddot{\chi}-3H\dot{\chi}-\sqrt{\frac{2}{3}}e^{-2\sqrt{\frac{2}{3}}\chi}
\left(\psi B_1(\phi,\psi)-\frac{1}{2}f_1(\phi,\psi) \right)\dot{\phi}^2+\frac{1}{2}\sqrt{\frac{2}{3}} e^{-\sqrt{\frac{2}{3}}\chi}\dot{\phi}\dot{\psi} -W_{,\chi}=0,
\end{equation}

\begin{equation}\nonumber
-e^{-\sqrt{\frac{2}{3}}\chi}\left(\psi B_1(\phi,\psi)-\frac{1}{2}f_1(\phi,\psi)\right)
\left(\ddot{\phi}+3H\dot{\phi}\right)+\frac{1}{2} \left(\ddot{\psi}+3H\dot{\psi}\right)- \frac{1}{2}\sqrt{\frac{2}{3}} \dot{\chi}\dot{\phi} -
\end{equation}

\begin{equation}\label{vph-1}
-\frac{1}{2}e^{-\sqrt{\frac{2}{3}}\chi}\left(\psi B_{1,\phi}-\frac{1}{2}f_{1,\phi}\right)\dot{\phi}^2-
e^{-\sqrt{\frac{2}{3}}\chi}\left(B_1+\psi B_{1,\psi}-\frac{1}{2}f_{1,\psi}\right)\dot{\phi}\dot{\psi}-
e^{\sqrt{\frac{2}{3}}\chi}W_{,\phi}=0,
\end{equation}

\begin{equation}\label{phi-1}
-\frac{1}{2}\left(\ddot{\phi}+3H\dot{\phi}\right)- \frac{1}{2}\sqrt{\frac{2}{3}}\dot{\chi}\dot{\phi} +
\frac{1}{2} e^{-\sqrt{\frac{2}{3}}\chi} \left(B_1+\psi B_{1,\psi}-\frac{1}{2}f_{1,\psi}\right)\dot{\phi}^2-
e^{\sqrt{\frac{2}{3}}\chi}W_{,\psi}=0.
\end{equation}

Einstein-Friedman equations are
\begin{equation}\label{EF-1}
3H^2=\gamma\left[\frac{1}{2} \dot{\chi}^2+ \frac{1}{2}e^{-2\sqrt{\frac{2}{3}}\chi}
\left(\psi B_1-\frac{1}{2}f_1 \right)\dot{\phi}^2 -\frac{1}{2}e^{-\sqrt{\frac{2}{3}}\chi}
\dot{\phi}\dot{\psi}\right]+W,
\end{equation}

\begin{equation}\label{EF-2}
\dot{H}=\gamma\left[-\frac{1}{2} \dot{\chi}^2- \frac{1}{2}e^{-2\sqrt{\frac{2}{3}}\chi}
\left(\psi B_1-\frac{1}{2}f_1 \right)\dot{\phi}^2 +\frac{1}{2}e^{-\sqrt{\frac{2}{3}}\chi}
\dot{\phi}\dot{\psi}\right].
\end{equation}

Here we include the parameter $\gamma$ to take into account phantom  zone of chiral fields.

\section{Zero potential}

Suggestion $W=0$, from $W=3H^2+\dot{H}$ gives the solution
\beq\label{H0}
H=\frac{1}{3}(t-t_*)^{-1},~t_*=const.,~~a(t)=a_0 t^{1/3},
\eeq
corresponding  for perfect fluid to  ultra-stiff matter.

Suggestion $W=0$ evidently restricts class of possible models, what can be reflected with parameter's function $B_1, B_2, f_1, f_2. $

The example of solution of chiral fields equations is
\beq
\phi=0,~~f_2(\phi,\psi)-\psi B_2(\phi,\psi)=0,~~\chi =c_\chi \ln (t-t_*) +\chi_*, ~~\psi=c_\psi \ln (t-t_*) +\psi_*,
\eeq
where $c_\chi, ~c_\psi,~\chi_*,~ \psi_*$ are some constants.
The solution above is valid only under relation
$f_2(\phi,\psi)=\psi B_2(\phi,\psi)$ for model's functions.
 Note that this relations leads to constraint on the model's form.
Others functions $B_1(\phi,\psi)$ and $f_1(\phi,\psi)$ have not any restrictions, they are free.

Let us set (additionally to $W=0$) $h_{22}=0$. I.e.
\beq
h_{22}=e^{-2\sqrt{\frac{2}{3}\chi}}\left( \psi B_1(\phi,\psi)-\frac{1}{2}f_1(\phi,\psi)\right)=0.
\eeq

 Thus the constrain on model's function is $B_1(\phi,\psi)=f_1(\phi,\psi)/2\psi$.

Dynamic equations \eqref{chi-1} - \eqref{phi-1} with $ \psi B_1(\phi,\psi)-\frac{1}{2}f_1(\phi,\psi)=0$ take the following view
\begin{equation}\label{chi-2}
-\ddot{\chi}-3H\dot{\chi}+\frac{1}{2}\sqrt{\frac{2}{3}} e^{-\sqrt{\frac{2}{3}}\chi}\dot{\phi}\dot{\psi} =0,
\end{equation}

\beq\label{psi-2}
\frac{1}{2} \left(\ddot{\psi}+3H\dot{\psi}\right)- \frac{1}{2}\sqrt{\frac{2}{3}} \dot{\chi}\dot{\phi}=0,
\eeq

\beq\label{phi-2}
-\frac{1}{2}\left(\ddot{\phi}+3H\dot{\phi}\right)- \frac{1}{2}\sqrt{\frac{2}{3}}\dot{\chi}\dot{\phi}=0.
\eeq

Note, that Hubble parameter $H(t)$ is known from  \eqref{H0}.

It is easy to see that subtracting \eqref{phi-2} from \eqref{psi-2} we get
\beq\label{ps-ph}
\left(\ddot{\psi}+3H\dot{\psi}\right)+ \left(\ddot{\phi}+3H\dot{\phi}\right)=0
\eeq

By setting
\beq\label{Q-1}
\left(\ddot{\psi}+3H\dot{\psi}\right)=Q(t),~~ \left(\ddot{\phi}+3H\dot{\phi}\right)=-Q(t)
\eeq
we make \eqref{ps-ph} as identical equality.

The solutions for \eqref{Q-1} are
\beq
\dot{\psi}=\frac{1}{t}\left(\int Q(t)tdt +C_\psi \right),
\eeq
\beq\label{sol-phi}
\dot{\phi}=\frac{1}{t}\left(-\int Q(t)tdt +C_\phi \right).
\eeq

We need no to integrate the equations above because in dynamic equations \eqref{chi-2} - \eqref{phi-2} we have derivatives of the fields $ \phi$ and  $\psi$ only.

Now we can define $\dot{\chi}$ from \eqref{phi-2}, substituting \eqref{sol-phi} and \eqref{Q-1} into \eqref{phi-2}. After some algebra we get
\beq
\dot{\chi}=\sqrt{\frac{3}{2}}Q(t)t \left( C_\phi -\int Q(t)tdt\right)^{-1}.
\eeq

Calculating the second derivative $\ddot{\chi}$
\beq
\ddot{\chi}=\dot{\chi}\left[ \frac{\dot{Q}}{Q}+\frac{1}{t} - \left( C_\phi -\int Q(t)tdt\right)^{-1}\right]
\eeq
we can insert the result into \eqref{chi-2}. The equation \eqref{chi-2} take the following view
\beq
\dot{\chi}\left[ \frac{\dot{Q}}{Q}+\frac{2}{t} - \left( C_\phi -\int Q(t)tdt\right)^{-1}\right]= \frac{1}{2}\sqrt{\frac{2}{3}}\frac{1}{t^2}e^{-\sqrt{\frac{2}{3}\chi}}
 \left(-\int Q(t)tdt +C_\phi \right)\left(\int Q(t)tdt +C_\psi \right).
\eeq

Thus we have ordinary differential equation with separable variables of the form
\beq
e^{\sqrt{\frac{2}{3}\chi}}d\chi=\frac{F_2(t)}{F_1(t)}dt,
\eeq
where
\beq
F_1(t)=\left[ \frac{\dot{Q}}{Q}+\frac{2}{t} - \left( C_\phi -\int Q(t)tdt\right)^{-1}\right],
\eeq
\beq
F_2(t)=\frac{1}{2}\sqrt{\frac{2}{3}}\frac{1}{t^2}
 \left(-\int Q(t)tdt +C_\phi \right)\left(\int Q(t)tdt +C_\psi \right).
\eeq

Finally we can define $\chi$ using the relation
\beq\label{fin}
e^{\sqrt{\frac{2}{3}\chi}}=\sqrt{\frac{2}{3}}\int \frac{F_2(t)}{F_1(t)}dt
\eeq

 Thus we can state that various chiral fields evolution may support evolution of the Universe with the scale factor \eqref{H0}.

Our next task is to show that the solutions space of \eqref{fin}  is not empty. To this end we may find few examples of solution by setting the function $Q(t)$ in various form.

Let us set $Q=\frac{C}{t}$, where $C= const.,~C_\phi=0,~C_\psi =0.$
Then the solution is
\beq
\chi = \sqrt{\frac{3}{2}}\ln \left|C_\chi-\frac{C^3}{4(C+1)}t^2\right|,~~C_\chi =const.
\eeq

Here are some examples of solutions when choosing the function $Q(t)$.

1. The function $Q(t)$ is a certain constant value: $Q = Q_{*}$.

\beq
\phi = -\frac{Q_{*}}{4}t^2 + c_{\phi}\ln|t| + C,
\eeq

\beq
\psi =\frac{Q_{*}}{4}t^2 + c_{\psi}\ln|t| + C,
\eeq

\beq
\chi = -\sqrt{\frac{2}{3}}\ln|Q_{*}t^2 - 2c_{\phi}|.
\eeq

2. The function $Q(t)$ depends on $t$ by power dependence: $Q = Q_{*}t^n$, where $n$ is any number.

\beq
\phi = \frac{Q_{*}t^{n+2}}{(n+2)^2} + c_{\phi}\ln|t| + C,
\eeq

\beq
\psi =-\frac{Q_{*}t^{n+2}}{(n+2)^2} + c_{\psi}\ln|t| + C,
\eeq

\beq
\chi = -\sqrt{\frac{2}{3}}\frac{\ln|Q_{*}t^{n+2} - c_{\phi}(n+2)|}{Q_{*}(n+2)} + C.
\eeq
Both examples show that the fields $\psi$ and $\phi$ can be linearly linked, and the field $\chi$ depends on $t$ logarithmically.

\section{Constant potential $W=W_*=const.$}

The solutions for gravitational field are well known, see for example, \cite{ChFCh20231}, \cite{ChFCh20232}.
The solution for Hubble parameter we find from the equation
\begin{equation}
3H^2+\dot{H}=W_*
\end{equation}
which is the sum of eqs. \eqref{EF-1} and \eqref{EF-2}.

The solutions are

1. Exponential rate of a scale factor

\begin{equation}
\label{EXP}
H=H_* =const,~~~a(t)=a_0 e^{H_* t},~~W_*=3H_*^{2},
\end{equation}
where $a_{0}$ is the scale factor at the beginning of inflation. 

2. Expansion defined by hyperbolic functions
\begin{equation}
\label{COSH}
H(t)=\sqrt{\frac{W_*}{3}}\tanh \left(\sqrt{3W_*}(t-t_*)\right),~~~~~a(t)= a_0\cosh^{1/3}\left(\sqrt{3W_*}(t-t_*)\right).
\end{equation}
where $\gamma =-1$, and

\begin{equation}
\label{SINH}
H(t)=\sqrt{\frac{W_*}{3}}\coth \left(\sqrt{3W_*}(t-t_*)\right),~~~ a(t)=a_0\sinh^{1/3}\left(\sqrt{3W_*}(t-t_*)\right).
\end{equation}
where $\gamma =1$.

3. Expansion defined by trigonometric functions

\begin{equation}
\label{COS}
H(t)=-\sqrt{\frac{W_*}{3}}\tan\left(\sqrt{3W_*}(t-t_*)\right),~~~~~ a(t)=a_0\cos^{1/3}\left(\sqrt{3W_*}(t-t_*)\right).
\end{equation}
where $\gamma =1$.

The solution (\ref{COS}) can be represented in another form:
\begin{equation}
\label{SIN}
H(t)=\sqrt{\frac{W_*}{3}}\cot\left(\sqrt{3W_*}(t-t_*)\right),~~~ a(t)=a_0\sin^{1/3}\left(\sqrt{3W_*}(t-t_*)\right).
\end{equation}
where $\gamma =1$.

Let us set the connection between model's function, as analog of the case above $W=0$,
\beq
f_2(\phi,\psi)=\psi B_2(\phi,\psi).
\eeq

Then we immediately get the relation between $\phi$ and $\chi$ in the form
\beq\label{phi-chi}
\phi=4W_*e^{\sqrt{\frac{2}{3}}\chi}.
\eeq

If we additionally set $h_{22}=0$ or $\psi B_1(\phi,\psi)-\frac{1}{2} f_1(\phi,\psi)=0$, then the dynamic equations \eqref{chi-2}-\eqref{phi-2} will be valid. The solutions of them can be represented in the general form leaving Hubble parameter included

\beq\label{Q-2}
\left(\ddot{\psi}+3H\dot{\psi}\right)=Q(t),
\eeq
\beq\label{Q-phi}
 \left(\ddot{\phi}+3H\dot{\phi}\right)=-Q(t).
\eeq

The solutions are
\beq
\dot{\psi}=\frac{1}{a^3(t)}\left(\int a^3(t)Q(t)dt +C_\psi \right),
\eeq
\beq\label{sol-phi}
\dot{\phi}=\frac{1}{a^3(t)}\left(-\int a^3(t)Q(t)dt +C_\phi \right).
\eeq

As we have $W=W_*=const.$ (and, consequently, derivatives on the fields will be equal to zero) then the dynamic equations \eqref{chi-2}-\eqref{phi-2} will not be changed. So we can apply the same approach as for the case with $W=0$.

Using \eqref{phi-chi} we can find connection between derivatives on $\phi$ with derivatives on $\chi$
\beq\label{dots}
\dot{\phi}= \sqrt{\frac{2}{3}}\phi \dot{\chi},~~
\ddot{\phi}= \sqrt{\frac{2}{3}}\phi \left( \ddot{\chi}+\sqrt{\frac{2}{3}}(\dot{\chi})^2\right).
\eeq

Using \eqref{dots} in \eqref{Q-phi} we can derive
\beq\label{Q-phi-H}
-Q(t)=\sqrt{\frac{2}{3}}\phi\left[\ddot{\chi} + \sqrt{\frac{2}{3}}(\dot{\chi})^2 + 3H\dot{\chi}\right].
\eeq

Substitution \eqref{Q-phi-H} into field equation \eqref{phi-2} gives
\beq
-\sqrt{\frac{2}{3}}\phi \left[\ddot{\chi} + 2\sqrt{\frac{2}{3}}(\dot{\chi})^2 + 3H\dot{\chi} \right]=0.
\eeq

Thus we can use the relation
\beq
\ddot{\chi}+3H\dot{\chi}=-2\sqrt{\frac{2}{3}}\dot{\chi}^2
\eeq
in \eqref{chi-2} to reduce the order of differential equation.

Equation \eqref{chi-2} takes the following view
\beq
2\sqrt{\frac{2}{3}}\dot{\chi}^2 + \frac{1}{2}\sqrt{\frac{2}{3}} e^{-\sqrt{\frac{2}{3}}\chi}\dot{\phi}\dot{\psi}=0.
\eeq

Using \eqref{phi-chi} and \eqref{dots}, and make some algebra we get
\beq
\dot{\chi}^2 +\sqrt{\frac{2}{3}}\dot{\chi}\dot{\psi}=0.
\eeq

So, we have the relation for $\dot{\chi}$ and $\dot{\psi}$ as follow
\beq
\dot{\chi}=-\sqrt{\frac{2}{3}}\dot{\psi}.
\eeq

Thus, we have the solution for chiral fields
\beq\label{phi-sol-g}
\phi= \int a(t)^{-3}\left[-F(t)+C_\phi\right]dt,
\eeq
\beq\label{psi-sol-g}
\psi= \int a(t)^{-3}\left[F(t)+C_\psi\right]dt,
\eeq
\beq\label{chi-sol-g}
\chi=- \sqrt{\frac{2}{3}}\int a(t)^{-3}\left[F(t)+C_\chi\right]dt,
\eeq
where
\beq\label{F-Q}
F(t)=\int a^3(t)Q(t)dt.
\eeq

Thus we have the solutions for gravitational field over Hubble parameter in the form \eqref{COSH}-\eqref{SIN}. To obtain the solutions for chiral fields it needs to insert each scale factor into solutions \eqref{phi-sol-g}-\eqref{chi-sol-g}.


Let us consider the following scale factor

\begin{equation}
a(t) = a_0\sinh^{\frac{1}{3}}\left(\sqrt{3W_{*}}\left(t - t_{*}\right)\right)
\end{equation}
as the example of solution.

For simplicity we will discard constants of integration.

1. $Q(t) = a_0^{-3}, (a_0=const.)$.
\begin{equation}
F(t) = \frac{\cos\left(\sqrt{3W_{*}}\left(t - t_{*}\right)\right)}{\sqrt{3W_{*}}},
\end{equation}

\begin{equation}
\phi = -\frac{\cos\left(\sqrt{3W_{*}}\left(t - t_{*}\right)\right)}{\sqrt{3W_{*}}},
\end{equation}

\begin{equation}
\psi = \frac{\cos\left(\sqrt{3W_{*}}\left(t - t_{*}\right)\right)}{\sqrt{3W_{*}}},
\end{equation}

\begin{equation}
\chi = -\sqrt{\frac{2}{3}}\frac{\cos\left(\sqrt{3W_{*}}\left(t - t_{*}\right)\right)}{\sqrt{3W_{*}}}.
\end{equation}

2. $Q(t) = \cosh\left(\sqrt{3W_{*}}\left(t - t_{*}\right)\right)$.
\begin{equation}
F(t) = \frac{a_0^3\sinh^2\left(\sqrt{3W_{*}}\left(t - t_{*}\right)\right)}{2\sqrt{3W_{*}}},
\end{equation}

\begin{equation}
\phi = -\frac{\cosh\left(\sqrt{3W_{*}}\left(t - t_{*}\right)\right)}{6W_{*}},
\end{equation}

\begin{equation}
\psi = \frac{\cosh\left(\sqrt{3W_{*}}\left(t - t_{*}\right)\right)}{6W_{*}},
\end{equation}

\begin{equation}
\chi = -\sqrt{\frac{2}{3}}\frac{\cosh\left(\sqrt{3W_{*}}\left(t - t_{*}\right)\right)}{6W_{*}}.
\end{equation}

3. $Q(t) = \frac{1}{a_0^3}\coth\left(\sqrt{3W_{*}}\left(t - t_{*}\right)\right)$.
\begin{equation}
F(t) = \frac{\sinh\left(\sqrt{3W_{*}}\left(t - t_{*}\right)\right)}{\sqrt{3W_{*}}},
\end{equation}

\begin{equation}
\phi = -\frac{t}{a_0^3\sqrt{3W_{*}}},
\end{equation}

\begin{equation}
\psi =\frac{t}{a_0^3\sqrt{3W_{*}}},
\end{equation}

\begin{equation}
\chi = -\sqrt{\frac{2}{3}}\frac{t}{a_0^3\sqrt{3W_{*}}}.
\end{equation}

Here are examples of solutions with the value of the scale factor:

\begin{equation}
a(t) = a_0\cosh^{\frac{1}{3}}\left(\sqrt{3W_{*}}\left(t - t_{*}\right)\right).
\end{equation}

1. $Q(t) = a_0^{-3}$.
\begin{equation}
F(t) = \frac{\sinh\left(\sqrt{3W_{*}}\left(t - t_{*}\right)\right)}{\sqrt{3W_{*}}},
\end{equation}

\begin{equation}
\phi = -\frac{a_0^3}{3W_{*}}\ln\left(\cosh\left(\sqrt{3W_{*}}\left(t - t_{*}\right)\right)\right),
\end{equation}

\begin{equation}
\psi = \frac{a_0^3}{3W_{*}}\ln\left(\cosh\left(\sqrt{3W_{*}}\left(t - t_{*}\right)\right)\right),
\end{equation}

\begin{equation}
\chi = -\sqrt{\frac{2}{3}}\frac{a_0^3}{3W_{*}}\ln\left(\cosh\left(\sqrt{3W_{*}}\left(t - t_{*}\right)\right)\right).
\end{equation}

2. $Q(t) = \sinh\left(\sqrt{3W_{*}}\left(t - t_{*}\right)\right)$.
\begin{equation}
F(t) = \frac{a_0^3\sinh^2\left(\sqrt{3W_{*}}\left(t - t_{*}\right)\right)}{2\sqrt{3W_{*}}},
\end{equation}

\begin{equation}
\phi = -\frac{1}{6W_{*}}\left[\sinh\left(\sqrt{3W_{*}}\left(t - t_{*}\right)\right)+\arctan\left(\sinh\left(\sqrt{3W_{*}}\left(t - t_{*}\right)\right)\right)\right],
\end{equation}

\begin{equation}
\psi = \frac{1}{6W_{*}}\left[\sinh\left(\sqrt{3W_{*}}\left(t - t_{*}\right)\right)+\arctan\left(\sinh\left(\sqrt{3W_{*}}\left(t - t_{*}\right)\right)\right)\right],
\end{equation}

\begin{equation}
\chi = -\sqrt{\frac{2}{3}}\frac{1}{6W_{*}}\left[\sinh\left(\sqrt{3W_{*}}\left(t - t_{*}\right)\right)+\arctan\left(\sinh\left(\sqrt{3W_{*}}\left(t - t_{*}\right)\right)\right)\right].
\end{equation}

3. $Q(t) = \frac{1}{a_0^3}\tanh\left(\sqrt{3W_{*}}\left(t - t_{*}\right)\right)$.
\begin{equation}
F(t) = \frac{\cosh\left(\sqrt{3W_{*}}\left(t - t_{*}\right)\right)}{\sqrt{3W_{*}}},
\end{equation}

\begin{equation}
\phi = -\frac{t}{a_0^3\sqrt{3W_{*}}},
\end{equation}

\begin{equation}
\psi =\frac{t}{a_0^3\sqrt{3W_{*}}},
\end{equation}

\begin{equation}
\chi = -\sqrt{\frac{2}{3}}\frac{t}{a_0^3\sqrt{3W_{*}}}.
\end{equation}
The examples given show that the scalar fields $\psi$, $\phi$, $\chi$ can be linearly related to each other.

 In this section once again we show that various evolution of scalar field may correspond to chosen scale factor.

\section*{Conclusion}

We considered in detail the way of transformation the model with higher derivations to GR with scalar fields proposed in \cite{Naruko:2015zze} with the aim to define the form of $f(R, (\nabla R)^2, \Box R)$ corresponding to chiral self-gravitating model. Special attention devoted to the model with reducing order, i.e. with the action \eqref{act-4-1}. This model under special choice of model's functions represented as the chiral self-gravitating model with the target space \eqref{ts3d} and the potential \eqref{V-1}. After that we study dynamic equations of the model in the cases with zero and constant potential. Examples of exact solutions show the linear connection between two fields and in few cases -- between all three fields.

Let us note about our intention to consider inverse task which gives  possibility to make transformation form obtained solutions for the chiral cosmological model in E-frame to the J-frame and to the solutions in modified $f(R, (\nabla R)^2,\Box R)$ gravity.

\section*{Acknowledgment}

The authors are grateful to K.A. Bronnikov for his interest in the work, useful discussions and his suggestion for expansion of the model.

The article was written within the framework of Additional Agreement No. 073-03-2024-
060/1 dated February 13, 2024 to the Agreement on the provision of subsidies from the federal
budget for financial support for the implementation of the state task for the provision of public
services (performance of work) No. 073-03-2024-060 dated January 18, 2024, concluded between
the Federal State Budgetary Educational Institution of Higher Education "UlSPU I.N. Ulyanov"
and the Ministry of Education of the Russian Federation.

\end{document}